\documentclass[aps,prl,floats,amsmath,showpacs,twocolumn]{revtex4}
\usepackage{graphicx}
\usepackage{multirow}

\begin{document}

\title{First-principles approach to lattice-mediated magnetoelectric
effects}

\author{Jorge \'I\~niguez}

\affiliation{Institut de Ci\`encia de Materials de Barcelona
(ICMAB-CSIC), Campus UAB, 08193 Bellaterra (Spain)}

\begin{abstract}
We present a first-principles scheme for the computation of the
magnetoelectric response of magnetic insulators. The method focuses on
the {\it lattice-mediated} part of the magnetic response to an
electric field, which we argue can be expected to be the dominant
contribution in materials displaying a strong magnetoelectric
coupling. We apply our method to Cr$_2$O$_3$, a relatively simple and
experimentally well studied magnetoelectric compound.
\end{abstract}


\maketitle

Magnetoelectric (ME) materials are insulators that allow control of
their magnetic properties by means of external electric
fields~\cite{fiebig05, eerenstein06}, thus attracting great
technological interest. Current research focuses on obtaining
compounds with a robust ME behavior at ambient conditions. This is
proving a major challenge, as progress is hampered by one fundamental
difficulty: the scarcity of ferromagnetic insulators (not to mention
ferromagnetic {\sl and} ferroelectric
multiferroics~\cite{filippetti02}) with a high Curie temperature. An
additional problem pertains to the magnitude of the effect: the ME
response is usually very small, reflecting the weakness of the
spin-orbit interactions that are typically responsible for the
coupling.

Quantum calculations based on efficient schemes like Density
Functional Theory (DFT) have proved very useful in studies of magnetic
and ferroelectric materials, and are expected to facilitate progress
on magnetoelectrics. Indeed, there is a growing number of DFT works
tackling the search for new compounds~\cite{ederer06} and even
proposing new coupling mechanisms~\cite{fennie06,picozzi07}. However,
we still lack a first-principles scheme to compute the ME coupling
coefficients, something that is critical to aid the experimental
work. In this Letter we introduce one such {\sl ab initio} methodology
and demonstrate its utility with an application to Cr$_2$O$_3$.

{\sl Lattice-mediated ME response.--} Computing the full ME response
of a material would require quantum relativistic simulations with
simultaneously applied electric and magnetic fields. While possible in
principle, such calculations constitute a great challenge even for the
case of static fields, and it is convenient to look for
simplifications of the problem.

In the following arguments we will consider an idealized
one-dimensional (1D) crystal displaying a linear ME effect, the
generalization being straightforward. At zero magnetic field, the
magnetization induced by the application of an electric field ${\cal
E}$ is given by:
\begin{equation}
\Delta {\cal M}({\cal E})\,=\,{\cal M}({\cal E})\,-\,{\cal
  M}^{S}\,=\,\alpha \, {\cal E}\,+\,{\cal O}({\cal E}^2)\, ,
\label{eq1}
\end{equation}
where $\alpha$ is the linear ME coefficient and we have included a
spontaneous magnetization ${\cal M}^{S}$ for generality. The magnitude
of the ME response is limited by the magnetic ($\chi^{\rm m}$) and
dielectric ($\chi^{\rm d}$) susceptibilities as $\alpha^2\, <
\,\chi^{\rm d} \chi^{\rm m}$~\cite{brown68}, which suggests that
strong ME couplings will occur in materials displaying large
dielectric and magnetic responses. On more physical grounds, one can
argue that large ME effects will be associated to significant
electronic hybridizations or orbital rearrangements induced by applied
electric fields, as it is processes of that nature that may lead to a
magnetic response via the spin-orbit coupling. It is then worth noting
that (1) such a response to an electric field is typical of
essentially all highly polarizable compounds used in applications and,
most importantly, (2) such strong dielectric responses are never a
purely electronic effect; rather, they are driven by the {\sl
structural} changes induced by the applied field. One can thus
conclude that large ME effects will most likely be based on {\sl
lattice-mediated} mechanisms.

Formally, the lattice-mediated contribution to the dielectric
susceptibility is defined as $\chi^{\rm d}_{\rm latt}\,=\,\chi^{\rm
d}\,-\,\chi^{\rm d}_{\rm elec}$, where $\chi^{\rm d}_{\rm elec}$
accounts for the purely electronic effect corresponding to clamped
atomic positions and lattice parameters. The ME coupling coefficient
$\alpha$ can also be decomposed in this way, and the discussion above
suggests that $\alpha_{\rm latt}$ will be the leading contribution in
materials displaying strong ME effects. We shall thus focus on its
computation.

{\sl Methodology.--} The structural response of an insulator to a {\sl
small} electric field can be modeled in terms of the infra-red (IR)
modes of the material, which are obtained from the diagonalization of
the force-constant matrix at the $\Gamma$ point of the Brillouin zone
(BZ). (Working with small fields allows us to truncate all the
following Taylor series at the lowest order possible.) Let us denote
by $u_i$ the amplitude of the $i$-th IR mode, with $i$ running from 1
to $N_{\rm IR}$, and by $C_i$ the corresponding eigenvalue. Taking the
$u_i$'s and the applied field ${\cal E}$ as independent variables, we
write the energy of our idealized 1D crystal around its equilibrium
state as
\begin{eqnarray}
\lefteqn{E(\{u_i \},{\cal E}) = \; E_0\,
+\frac{1}{2}\sum_{i=1}^{N_{\rm IR}}C_i u_i^2 } \nonumber \\
& & - \, \Omega_0 \, {\cal E} \, [ {\cal P}^{S} \, + \,
    \chi^{\rm d}_{\rm elec}\,{\cal E} \, + \, \Delta {\cal P}_{\rm
    latt}(\{u_i\}) ] \, ,
\label{eq2}
\end{eqnarray}
where $\Omega_0$ is the unit cell volume and we have included a
spontaneous polarization ${\cal P}^{S}$ for generality. Note we have
decomposed the linear part of the induced polarization into electronic
($\chi^{\rm d}_{\rm elec}\,{\cal E}$) and lattice contributions, the
latter being given by
\begin{equation}
\Delta {\cal P}_{\rm latt}\,=\,\frac{1}{\Omega_0}\sum_{i=1}^{N_{\rm
IR}} p^{\rm d}_i u_i \, ,
\label{eq3}
\end{equation}
where $p^{\rm d}_i$ is the polarity of the $i$-th IR mode, which can
be obtained from the atomic Born effective charges and the mode
eigenvector~\cite{b-ghosez06}. From these expressions, the equilibrium
value of $u_i$ for a given ${\cal E}$ is calculated to be
\begin{equation}
u_i\,=\,\frac{p^{\rm d}_i}{C_i}{\cal E}\, .
\label{eq4}
\end{equation}
On the other hand, the assumption that our model crystal displays a
linear ME effect implies
\begin{equation}
\Delta {\cal M}(\{ u_i \},{\cal E})\,=\,\alpha_{\rm elec}\, {\cal
  E}\,+\,\frac{1}{\Omega_0}\sum_{i=1}^{N_{\rm IR}}p^{\rm m}_i u_i \, ,
\label{eq5}
\end{equation}
where the $u_i$'s are again assumed to be independent of ${\cal E}$
and we have introduced $p^{\rm m}_i$ parameters that quantify the
magnetic response to the IR modes. Finally, by virtue of
Eq.~(\ref{eq4}) we can write
\begin{equation}
\Delta {\cal M}({\cal E})\,=\,\alpha_{\rm elec}\, {\cal E}\,+\,
\frac{1}{\Omega_0}\sum_{i=1}^{N_{\rm IR}} \frac{p^{\rm m}_i p^{\rm
d}_i}{C_i} {\cal E}\, ,
\label{eq6}
\end{equation}
and it immediately follows that
\begin{equation}
\alpha_{\rm latt}\,=\,\sum_{i=1}^{N_{\rm IR}} \alpha_{{\rm latt},i}
\,=\, \frac{1}{\Omega_0}\sum_{i=1}^{N_{\rm IR}} \frac{p^{\rm m}_i
p^{\rm d}_i}{C_i}\, ,
\label{eq7}
\end{equation}
where mode-dependent contributions to $\alpha_{\rm latt}$ have been
defined. This equation encapsulates our method for an {\sl ab initio}
computation of the ME response. Its most remarkable feature is that
all the parameters that appear in it can be computed without the need
of simulating the material under applied electric or magnetic fields,
which brings the calculation of ME effects within the scope of the
most widely used DFT codes.

The above expression offers some insight into the microscopic
ingredients needed to have a strong lattice-mediated ME response. In
essence, one would like to have {\sl soft} IR modes (i.e., with small
$C_i$) that are highly polarizable (i.e., with large $p^{\rm d}_i$)
and cause a large magnetic response (i.e., with large $p^{\rm
m}_i$). Note that the $p^{\rm m}_i$ parameters can be viewed as the
magnetic analogue of the polarities $p^{\rm d}_i$ associated to the
dynamical charges. Also in analogy with $p^{\rm d}_i$, $p^{\rm m}_i$
can be written as a sum of atomic contributions weighted by the
corresponding components of the IR eigenvector. It is then clear that,
in order to have IR modes with simultaneously large $p^{\rm d}_i$ and
$p^{\rm m}_i$, we need materials in which the magnetic atoms present
large Born effective charges. While rare, this is the case of
compounds like CaMnO$_3$~\cite{filippetti02}.

A few additional comments are in order. (1) The proposed approach can
be used to simulate ME effects of arbitrary order in ${\cal E}$, but
is restricted to couplings that are linear in the magnetic field. This
limitation can be remedied by simulating the material under applied
magnetic fields, which is relatively easy in comparison with treating
finite electric fields in extended systems described with periodic
boundary conditions~\cite{souza02}. (2) The ME response mediated by
the strain ($\eta$) can be trivially included. More specifically, the
above formulas are correct for paraelectrics, for which the leading
strain terms in Eq.~(\ref{eq2}) are of the form $\eta^2$ and $\eta
u_i^2$, and result in ME responses of order higher than linear. In
ferroelectrics, on the other hand, there exist $\eta u_i$ terms that
give a linear contribution to the response. (3) While the above
derivation is made in terms of the eigenvectors of the force-constant
matrix, one could imagine an analogous scheme using the IR eigenmodes
of the dynamical matrix as structural variables. It would then be
possible to model the dynamical ME response. (4) It is possible to
derive the above results in a more general way, by working with an
energy $E(\{u_i \},{\cal E}; \{m_j \},{\cal H})$ that includes the
localized magnetic moments $m_j$ and the magnetic field ${\cal H}$ as
independent variables of the system.

\begin{figure}
\includegraphics[width=0.38\columnwidth]{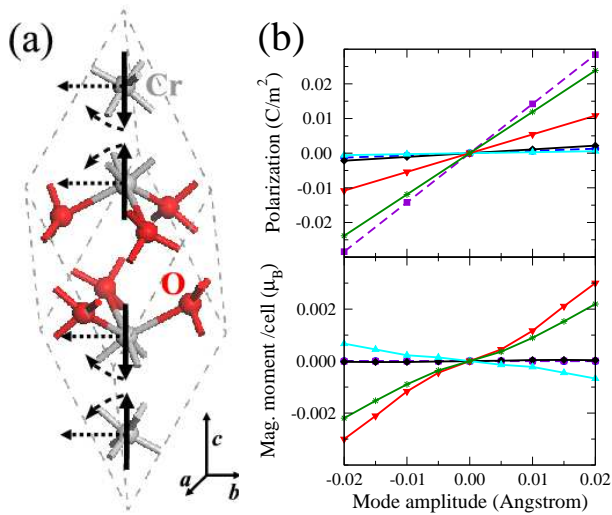}
\includegraphics[width=0.55\columnwidth]{fig1b.eps}
\caption{Panel~a: Primitive cell of Cr$_2$O$_3$. Solid arrows
  represent the AFM ground state. Dashed arrows sketch the atomic
  displacements within the $ab$ plane associated to a typical $E_u$ IR
  mode, as well as the induced spin rotations that render a net
  magnetic moment. Panel~b: Computed polarization and magnetization
  induced by the condensation of the IR modes. Dashed and solid lines
  correspond to $A_{2u}$ and $E_u$ modes, respectively. Note that the
  polarizations and magnetizations associated to the $E_u$ ($A_{2u}$)
  modes lie within the $ab$ plane (along the $c$ direction). Note also
  that the magnetization induced by the $A_{2u}$ modes is essentially
  zero.}
\label{fig1}
\end{figure}

{\sl Results for Cr$_2$O$_3$.--} The work on magnetoelectrics starts
with the prediction~\cite{dzyaloshinskii59} and experimental
confirmation~\cite{astrov6061,rado-exp} that linear ME effects occur
in Cr$_2$O$_3$ (chromia), which remains one of the simplest and best
studied ME materials. Cr$_2$O$_3$ is an antiferromagnetic (AFM)
insulator with a 10-atom unit cell and the magnetic structure sketched
in Fig.~\ref{fig1}. The magnetic easy axis lies along the rhombohedral
direction $c$. The crystal has the magnetic space group $R\bar{3}'c'$,
which preserves all the crystallographic symmetries of $R\bar{3}c$;
thus, the compound is paraelectric. Cr$_2$O$_3$ presents six IR modes:
two polarized along the rhombohedral $c$-axis, corresponding to the
$A_{2u}$ irreducible representation of $3m$, and four
double-degenerate modes with $E_u$ symmetry and polarization within
the $ab$-plane. The linear ME tensor is diagonal with two independent
terms: $\alpha_{aa}=\alpha_{bb}=\alpha_{\perp}$ and
$\alpha_{cc}=\alpha_{\parallel}$. Naturally, the lattice-mediated part
of $\alpha_{\perp}$ ($\alpha_{\parallel}$) can be decomposed into
contributions from the $E_u$ ($A_{2u}$) modes, which we can compute
with our method. (In the following we drop the ``latt'' subscript from
the $\alpha$'s to alleviate the notation.)

For the calculations we used the LDA~\cite{lda} approximation to DFT
as implemented in the plane-wave code VASP~\cite{vasp}. We used the
PAW scheme~\cite{paw} to represent the atomic cores. Only the nominal
valence electrons were explicitly solved, which we checked is
sufficient. Let us just note that all the {\sl trivial} calculations
involved in this study (e.g., for the equilibrium atomic structure,
force-constant matrix, or induced polarizations~\cite{kingsmith93})
were performed accurately and following well-established procedures,
and that all of them were done at the collinear level. To obtain the
$p^{\rm m}_i$ parameters in Eq.~(\ref{eq7}), we computed the magnetic
response upon condensation of the IR modes by running fully
self-consistent non-collinear simulations including spin-orbit
couplings. Interestingly, we found that a non-self-consistent
approach, as usually employed for the computation of magnetic
anisotropy energies, renders qualitatively incorrect results in this
case. Let us also stress that, given the small magnitude of the energy
differences associated to the ME effects in Cr$_2$O$_3$, one has to be
very careful with the choice of the parameters controlling the
accuracy of the calculations. In particular, we found it necessary to
use a very demanding stopping criterion for the self-consistent-field
calculations (namely, energies converged down to 10$^{-10}$~eV) to
obtain, in a computationally robust way, reliable values of the
magnetic moments induced by the condensation of the IR modes. We also
determined that a $k$-point grid of at least 7$\times$7$\times$7 is
needed for accurate BZ integrations. (A magnetic easy axis in the
$ab$-plane is incorrectly predicted if grids that are not dense enough
are used.)  The plane-wave cutoff was found to be less critical; we
used 400~eV. We employed the ``LDA+U'' scheme of Dudarev {\it et
al.}~\cite{dudarev98} for a better treatment of the 3$d$ electrons of
Cr. We chose $U_{\rm eff}=2$~eV, which renders results in acceptable
agreement with experiment for the atomic structure, IR phonon
frequencies, electronic band gap, and magnetic
moments~\cite{fn-trivialresults}. At any rate, we checked the choice
of $U_{\rm eff}$ is not critical, even for the computation of ME
coefficients. Finally, let us note the orbital degrees of freedom can
be expected to be {\sl quenched} in Cr$_2$O$_3$; thus, we neglected
their contribution to the magnetization.

\begin{table}
\caption{Parameters of Eq.~(\protect\ref{eq7}) computed for the IR
  modes of Cr$_2$O$_3$. Modes are divided in two groups, $A_{2u}$ and
  $E_u$, according to their symmetry. The last line shows the results
  for the two independent $\alpha$ coefficients, obtained from the
  addition of the corresponding mode contributions. The $\alpha$'s are
  given in Gaussian units (g.u.)~\protect\cite{rivera94}.}
\vskip 2mm
\begin{tabular*}{0.95\columnwidth}{@{\extracolsep{\fill}}lrr@{\hspace{4mm}}r@{}rrr}
\hline\hline
 & \multicolumn{2}{l}{$A_{2u}$ modes} & \multicolumn{4}{c}{$E_u$
 modes} \\
\hline
$C_i$ (eV/\AA$^2$) & 10.8 & 25.7 & 10.4 & 16.9 & 21.6 & 32.5 \\
$p^{\rm d}_i$ ($|e|$)  & 0.39 & 8.52 & 0.65 & 0.16 & 3.24 & 7.14 \\
$p^{\rm m}_i$ (10$^{-2}\mu_{\rm B}$/\AA) & 0.02 & 0.04 & 0.41 & $-$2.70 & 11.32 &
8.51 \\
$\alpha_i$ (10$^{-4}$  g.u.) & 0.00 & 0.00 & 0.01 & $-$0.01 &
 0.62 & 0.68 \\
\hline
\multicolumn{1}{c}{$\sum_i \alpha_i$ (10$^{-4}$  g.u.)} &
 \multicolumn{2}{l}{$\alpha_{\parallel}$= 0.00} &  
\multicolumn{4}{c}{$\alpha_{\perp}$= 1.30} \\ 
\hline\hline
\end{tabular*}
\label{tab1}
\end{table}

Table~\ref{tab1} and Fig.~\ref{fig1} summarize our results, which
present the following features. (1) We obtain $\alpha_{\parallel}$
much smaller than $\alpha_{\perp}$. Indeed, our calculations indicate
that the magnetic response associated to the $A_{2u}$ modes is nearly
zero, and provide an explanation for such an effect. We find that, for
the $E_u$ modes, the induced in-plane magnetization occurs via a
canting of the Cr spins, as sketched in Fig.~\ref{fig1}. In contrast,
in the case of the $A_{2u}$ modes, no symmetry-allowed spin canting
can induce a magnetization along the $c$ direction. Instead, the
simulations show that the magnetization originates from a tiny charge
transfer between the spin-up and spin-down Cr sublattices. Probably,
the smallness of the corresponding $p^{\rm m}_i$ coefficients reflects
the relatively large energy cost associated to such a mechanism. (2)
The ME response $\alpha_{\perp}$ is dominated by the hardest $E_u$
modes and, interestingly, such a result could have been anticipated
from the mode eigenvectors. More precisely, the two hardest modes
present a relatively large Cr contribution, which should lead to
relatively large values of $p^{\rm m}_i$, as we indeed find. In
addition, in the hardest eigenmode the Cr and O sublattices move
rigidly and in opposite directions, which must result in a large
$p^{\rm d}_i$, exactly as found. (3) We obtain both positive (from
three modes) and negative (from one mode) contributions to
$\alpha_{\perp}$. (Given the smallness of the magnetic effects
computed, we have not tried to identify the electronic underpinnings
of having positive or negative $\alpha_i$'s.) This result suggests
that, in a general case, a small static ME effect may be the result of
cancellations between contributions from different IR modes. Hence,
large static ME effects will most likely be associated to compounds in
which a single IR mode dominates the response.

To the best of our knowledge, the low temperature ME response of
Cr$_2$O$_3$ is not totally understood, which reflects both the
difficulties involved in ME measurements and the rich nature of the
problem. The experimental results at 4.2~K are quite
scattered~\cite{wiegelmann94}: $|\alpha_{\perp}|$ ranges from
0.2$\times$10$^{-4}$ to 4.7$\times$10$^{-4}$~in Gaussian units (g.u.)
and $|\alpha_{\parallel}|$ from 0.4$\times$10$^{-4}$ to
1.2$\times$10$^{-4}$~g.u. There are reasons to believe that the
magnitude of the ME effects was underestimated in the early
experiments~\cite{fn-MEerrors}, and that the largest coefficients
measured~\cite{wiegelmann94,kita79} are the most reliable ones. In
particular, $|\alpha_{\perp}|$ probably lies somewhere between
2$\times$10$^{-4}$ and 4$\times$10$^{-4}$~g.u., which is remarkably
close to our result. Interestingly, it is not clear how to explain
this relatively large value of $\alpha_{\perp}$ in terms of the purely
electronic mechanisms typically
considered~\cite{rado-theo,hornreich67,kita79}. It is thus worth
noting our computed lattice-mediated ME response is of the same
magnitude as the one measured. As for the parallel response, all the
experiments render $|\alpha_{\parallel}|\,<\,|\alpha_{\perp}|$ at low
temperatures, but none reports an essentially zero value as we
obtain. Our results are thus compatible with the notion that either a
purely electronic mechanism, as the electric-field-induced $g$ shift
proposed in Ref.~\cite{hornreich67}, or a magnetic effect not related
to the ME coupling~\cite{rado-theo} is responsible for the non-zero
$\alpha_{\parallel}$ at low temperatures.

In summary, we have introduced an efficient method to compute
lattice-mediated magnetoelectric responses {\sl ab initio}. We hope
our work will enable a more effective interaction between theory and
experiment in the search for materials that can be used in
applications.

Very fruitful discussions with Ph.~Ghosez are gratefully
acknowledged. This is work funded by MaCoMuFi (STREP\_FP6-03321). It
also received some support from CSIC (PIE-200760I015), the Spanish
(FIS2006-12117-C04-01, CSD2007-00041) and Catalan (SGR2005-683)
Governments, and FAME-NoE. Use was made of the Barcelona
Supercomputing Center (BSC-CNS).


\begin{thebibliography}{99}
\bibitem{fn-trivialresults} We worked at the experimental volume
  (96~\AA$^3$); selected quantities were computed at the theoretical
  volume (92~\AA$^3$), and no significant difference was
  observed. Rhombohedral unit cell: $a$~=~5.37~\AA\ and
  $\alpha$~=~54.66$^{\circ}$; symmetry independent atoms: Cr at
  ($x$,$x$,$x$) with $x$~=~0.1536 and O at (0.9425, 0.5575, $1/4$);
  band gap: $\sim$~2.1~eV; magnetic moment per Cr:
  $\sim$~2.8~$\mu_{\rm B}$. IR mode frequencies (cm$^{-1}$): $A_{2u}$
  (408, 597) and $E_u$ (316, 455, 578, 653). These results are in
  reasonable agreement with experimental data and previous
  calculations reported in Refs.~\protect\onlinecite{lucovsky77}
  and~\protect\onlinecite{rohrbach04}.
\bibitem{fn-MEerrors} A major difficulty is related with the fact that
different AFM domains may present ME effects of opposite
sign~\protect\cite{astrov6061,rado-exp}. Hence, the magnitude of
$\alpha$ measured in a sample in a multi-domain state will be smaller
than that of the actual {\sl intrinsic} response.
\end{thebibliography}
\end{document}